\documentstyle[aps,prl,graphicx,floats]{revtex}
\begin{document}
\draft

\twocolumn[\hsize\textwidth\columnwidth\hsize\csname @twocolumnfalse\endcsname

\title{ Structural glass on a lattice in the limit of infinite dimensions.}

\author{A.V. Lopatin$^{1,2}$  and L.B. Ioffe$^{1}$ }
\address{$^{1}$Department of Physics, Rutgers University, Piscataway,
New Jersey  08854 \\
$^{2}$ Department of Physics, University of Maryland, College Park, 
MD 20742-4111}

\date{\today}
\maketitle

\begin{abstract}
We construct a mean field theory for the lattice model of a structural glass 
and solve it using the replica method and one step replica symmetry breaking 
ansatz; this theory becomes exact in the limit of infinite dimensions. 
Analyzing stability of this solution we conclude that the metastable states 
remain uncorrelated in a finite temperature range below the transition, but 
become correlated at sufficiently low temperature. We find dynamic and 
thermodynamic transition temperatures as functions of the density and 
construct a full thermodynamic description of a typical physical process 
in which the system gets trapped in one metastable state when cooled below 
vitrification temperature. We find that for such physical process the entropy 
and pressure at the glass transition are continuous across the transition 
while their temperature derivatives have jumps.

\end{abstract}

\vskip2pc]

\bigskip

Thermodynamics of structural glasses and of vitrification transition is a
long standing problem in statistical physics. The essential feature of glass
formation is a division of the phase space into an exponentially large number of
similar compartments. The system gets trapped in one of these compartments
and stays there forever in a complete defiance of ergodic hypothesis; this
defines dynamical ''states'' of the system. Each of these states is
characterized by a local order parameter (e.g. atoms positions,
magnetization) that varies in space and distinguishes one state from
another. This makes even a mean field theory difficult to construct because
the thermodynamic ground state of the system (or any other particular
state) has a unique configuration of the order parameter field, thus making
it difficult to describe system in terms of average (and site independent)
quantities. Few ways to resolve this difficulty were proposed recently 
\cite{Marinari94,Chandra95,Monasson95,Franz97,Mezard99a,Mezard99b,Ioffe2001}; 
they all share a common idea that states of the system with the same energy 
per site (but perhaps different total energies) are essentially equivalent and
instead of studying one particular state one can average over all states
with the same energy density.

There are a few ways to implement averaging over states, the most convenient
ones seem to be the clone method \cite{Mezard99a,Mezard99b} and the introduction
of a small random field conjugated to the order parameter (magnetization in
case of spin glasses or density in structural glasses) \cite{Ioffe2001}. The
physical idea of the latter is that an infinitesimally small random field
applied to a system with many metastable states rearranges the energies of
low-lying states making the problem similar to one with quenched disorder.
In a spin system, for example, we add to the Hamiltonian a magnetic field
part: ${\cal H}\rightarrow {\cal H}+\sum_{i}h_{i}S_{i}$ with small random $%
h_{i}$. The resulting change in the energy of a typical metastable state is
of the order of $\sqrt{N}h$; because this energy interval contains a large
number of metastable states, we expect that a small non-zero field would
result in a large rearrangement of their energies but would not change the
properties of individual states. Averaging over the random field
configurations may be then performed in a usual way introducing $n$ replicas
of the system and taking the limit $n\rightarrow 0$. The alternative idea of
the clone method is to consider a system of $m$ clones constrained to be in
the same metastable states by an infinitesimally weak coupling. Generally,
the free energy density of the clone system may be written as 
$F_{clone}=-TN^{-1} \ln \int
dFe^{-N(m\beta F-S_{conf}(F))}$ where $S_{conf}(F)$ is configurational
entropy ($S_{conf}=\frac{1}{N}\ln ({\cal N}_{states})$). If
$S_{conf}(F)$ is concave  and $dS_{conf}(F)/dF$
is finite at the lower bound $F=F_{0}$ (corresponding to the
ground state) the appropriate choice of $m$ ($m\propto T$ at low $T$) leads
to a partition function dominated by a small vicinity of any given $F>F_{0}$
that still contains thermodynamic number of states providing the averaging
mechanism in this approach. In this paper we shall apply the mean field
formalism developed in our paper \cite{Ioffe2001} that combines random field
approach and the locator expansion developed for the glass physics in \cite
{Feigelman} to the simplest model of a realistic structural glass. The mean
field theory for the structural glasses can be justified formally only in
the limit of infinite dimensions;  thus, in our derivation of the mean field
theory below we shall consider the space of arbitrary dimension, $d$, and
keep only the leading terms in $1/d$. We hope, however, that it provides
reasonable results even for $d=3$ because the actual parameter is the
coordination number which is already large in three dimensions. 

Another important physical assumption that simplifies the calculations
significantly is that different states are uncorrelated. Formally, this
allows one to look for one step replica symmetry breaking (RSB) solution of
the mean field equations. We check the stability of this solution below and
find that it is stable in some temperature range below the glass transition
temperature but that it becomes unstable at low temperatures. The results of
one step RSB  can be translated much easier into the physical properties
than results of a full RSB because in this case one can establish a formal
equivalence between this method and a clone approach with the number of
clones being equal to the size of the blocks in 1RSB ansatz $m,$ and replica
free energy $F_{repl\text{ }}$ being equal to the free energy of the clone
system per one clone $F_{repl}=F_{clone}/m.$   For a given $m$ the main
contribution to $F_{clone}$ comes from states with  $m=T dS_{conf}(F)/dF$ , 
thus 
\begin{equation}
S_{conf}=\beta m^{2}\frac{\partial }{\partial m}(F_{clone}/m),\,\,\,F=\frac{%
\partial }{\partial m}F_{clone.}  \label{sclon}
\end{equation}

Metastable states first appear at the dynamical transition temperature $%
T_{g},$ at which the system becomes trapped in one of the metastable states
with largest possible free energy $F$ because these states dominate the
configuration space since $S_{conf}(F)$ is a monotonically increasing
function. Under a further decrease of temperature the system remains trapped
in the same metastable state, {\it thus the thermodynamics of a physical
cooling process is determined by the free energy of a single metastable
state.} To describe this physical process by a replica  method we need to
find the dependence $m(T)$ consistent with the requirement that the physical
system is trapped in a single metastable state. We assume that all states
that were similar at one temperature remain similar as the temperature is
decreased and these states do not bifurcate or disappear (consistent with
one step RSB);  this implies that configuration entropy $S_{conf}$ is
constant along the trajectory $m(T)$ corresponding to a physical cooling
process. We check that this assumption is consistent with (\ref{sclon}) that
gives the free energy of a metastable state and the corresponding
configuration entropy in terms of the replica free energy. First, we note
that  the usual thermodynamic identity $E=\partial \lbrack \beta
F_{repl}(\beta ,m)]/\partial \beta $ that relates internal energy to the
free energy remains valid in replica approach. Second, if the system is
trapped in a single metastable state, the thermodynamic identity $E(\beta
,m)=d[\beta F(\beta ,m)]/d\beta $ must also hold.  Remarkably, these
equations  are consistent with each other and with (\ref{sclon}){\it \ if
and only if  }$S_{conf}(\beta ,m)=const$. This gives us the implicit
equation for $m(T)$ and allows us to deduce the thermodynamics of the system
trapped in a single metastable state provided \ one step RSB is a stable
solution. 

We now turn to the details of the model and its solution. We consider the
simplest glass model which contains only short range repulsion between
atoms:  

\begin{equation}
H={\frac{1}{2}}\sum_{i.j}\rho _{i}J_{i,j}\rho _{j}  \label{ham}
\end{equation}
where $i,j$ are the sites of a d-dimensional lattice and $\rho _{i}=1,0$
represents the on-site density. The glass is formed for such coupling
matrices $J_{i,j}$ that do not allow low energy crystal states (density
waves). It is convenient to represent  $J_{i,j}$ in the form $\hat{J}=uf(%
\hat{t}/\sqrt{2d}),$ where  $\hat{t}$ is the nearest neighbor hopping
operator and we introduced the scaling $\sqrt{2d}$ that is needed to get a 
sensible limit at $d\rightarrow \infty \cite{Georges}.$ The simplest choice
of the function $f(\varepsilon )\sim \varepsilon $ leads to the crystal
ground state with a chess-board ordering for density $\bar{\rho} 
\approx 1/2.$ We shall take $f(\varepsilon )=\varepsilon ^{2}$ and show that
this frustrated interaction leads to the formation of the glass state at low
temperature.

The Hamiltonian (\ref{ham}) may be represented as $H=H_{0}+H^{\prime },$
separating out large but physically irrelevant constant term $H_{0}=ud
\bar{\rho }^{2}$. We focus on the nontrivial term $H^{\prime
}=\sum_{i,.j}\tilde{\rho}_{i}J_{i,j}\tilde{\rho}_{j}$ where $\tilde{\rho}%
=\rho -\bar{\rho}.$ Introducing the replicas we get

\begin{equation}
\beta H^{\prime }=\frac{\beta }{2}\sum_{i,j,\alpha }\tilde{\rho}_{i}^{\alpha
}J_{i,j}\tilde{\rho}_{j}^{\alpha }-\sum_{i,\alpha }\mu ^{\alpha }\tilde{\rho}%
_{i}^{\alpha }  \label{hamp}
\end{equation}
where $\alpha $ is the replica index taking values form $0$ to $n,$
and $\mu _{\alpha }$ are the the Lagrange multipliers (equal to the chemical
potential divided by the temperature at the saddle point) that impose 
constraint on the number of
particles in each replica. In the limit of infinite dimensions the original
model (\ref{hamp}) can be replaced with an effective single site model

\begin{equation}
-\beta H_{eff}^{\prime }=\frac{1}{2}U({\cal B})-\frac{1}{2}\tilde{\rho }%
_{\alpha }B_{\alpha ,\beta }\tilde{\rho }_{\beta }+\mu _{\alpha }%
\tilde{\rho }_{\alpha }  \label{heff}
\end{equation}
where ${\cal B}$ is the matrix in replica space and $\tilde{\rho}=\rho -\bar{%
\rho}$ has only replica index. The mean field self-consistency condition is
that $B$ gives the saddle point of the free energy: $\langle \tilde{\rho 
}_{\alpha }\tilde{\rho }_{\beta }\rangle =dU({\cal B})/dB_{\alpha \beta }
$. The potential $U({\cal B})$ can be determined from the condition that all
single site correlation functions of the model (\ref{heff}) coincide with
the correlation functions of the original model (\ref{hamp}). Instead of
comparing these correlation functions directly we decouple the interaction
term introducing the auxiliary fields, $v^{\alpha }$, conjugated to $\tilde{%
\rho}$  
\begin{eqnarray}
-\beta H^{\prime }=&-&\frac{T}{2}\sum_{i,j,\alpha }v_{i}^{\alpha
}J_{i,j}^{-1}v_{j}^{\alpha }+\sum_{i,\alpha }\tilde{\rho}_{i}^{\alpha
}(iv_{i}^{\alpha }+\mu ^{\alpha })
\nonumber \\
&-&\frac{n}{2}{\rm Tr}\log \hat{J}\beta 
\label{hampcon} \\
-\beta H_{eff}^{\prime }=&-&\frac{1}{2}\sum_{\alpha ,\beta }v^{\alpha
}B_{\alpha ,\beta }^{-1}v^{\beta }+\sum_{\alpha }\tilde{\rho}^{\alpha
}(iv^{\alpha }+\mu ^{\alpha })
\nonumber \\
&+&\frac{1}{2}U(B)-\frac{1}{2}{\rm Tr} \log B  \label{heffcon}
\end{eqnarray}
and compare the correlation functions of these fields. Summation over $%
\tilde{\rho}$ in both models results in the on-site interaction of the
fields $v.$ Inspecting the terms of the perturbation theory in this on-site
interaction for the correlator $G_{i,j}^{\alpha \beta }=\langle
v_{i}^{\alpha }v_{j}^{\beta }\rangle $ one verifies that in the leading
order in $1/d$ it is given by $\hat{G}=[T\hat{J}^{-1}-\Sigma ]^{-1}$ with
the self energy $\Sigma $ which is diagonal in the site index: $\Sigma =-(%
{\cal A}\,\beta u)^{-1}\delta _{ij}$. This approach is similar to a locator
expansion \cite{Feigelman} but in our case the locator ${\cal A}$ is
non-trivial in the replica space. The single site correlation function $%
{\cal G}_{\alpha \beta }\equiv G_{i,i}^{\alpha \beta }$ (that we need to
establish the correspondence between the models) is 
\begin{equation}
{\cal G}=u\beta \int d\epsilon \nu (\epsilon )[1/f(\varepsilon )+{\cal A}%
^{-1}]^{-1}  \label{G}
\end{equation}
where $\nu (\varepsilon )=\exp [-\varepsilon ^{2}/2]/\sqrt{2\pi }$ is the
density of states in the limit $d\rightarrow \infty .$

Now we turn to the model (\ref{heffcon}). Here the self energy is diagonal
in the site index by construction, further, the interaction part of this
model is the same as for model (\ref{hampcon}); assuming that their single
site correlation functions coincide, we get that their single site
self-energies are equal, so that for this model  
\begin{equation}
{\cal G}=[{\cal B}^{-1}+(\beta u{\cal A)}^{-1}]^{-1}  \label{G1}
\end{equation}
which gives us equation relating ${\cal B}$ and ${\cal A}$. Further, the
correlation function of the auxiliary fields can be expressed through ${\cal D%
}=\langle \tilde{\rho }_{\alpha }\tilde{\rho }_{\beta }\rangle $ via 
${\cal G=B-B}{\cal D}{\cal B}$, combining this equation with the saddle
point condition and using $\int d\varepsilon D(\varepsilon )=1$ we find the
implicit expression for the potential 

\[
U(B)={\rm Tr}
\ln [{\cal A}+{\cal B}/u\beta ]-{\rm Tr}
\int d\epsilon D(\epsilon )\ln (
{\cal A}+f(\varepsilon ))
\]
that has to be evaluated at the saddle point with respect to ${\cal A}.$ 

{\it Liquid state. }In the liquid phase we take the replica symmetric ansatz 
$A_{\alpha ,\beta }=a\delta _{\alpha ,\beta },B=b\delta _{\alpha ,\beta }$,
exclude the chemical potential $\mu $ and get 
\begin{equation}
\beta F=[L(a)-\ln (a+b/u\beta )+b\bar{\rho }(1-\bar{\rho})]/2-S_{0},
\label{fliq}
\end{equation}
where $L(a)=\int d\varepsilon \nu (\varepsilon )\ln (a+\varepsilon ^{2})$
and $S_{0}=-\bar{\rho}\ln \bar{\rho}-(1-\bar{\rho})\ln (1-\bar{\rho})$ is
the high temperature entropy. The corresponding energy is $E=T\bar{\rho}(1-%
\bar{\rho})b/2.$ Numerical solution of the saddle point equations show that
at any density the entropy becomes negative at low  temperatures. The
dependence of the temperature at which the entropy becomes zero $T_{S=0}$ on
the density for the coupling constant $u=1$ is shown in Fig. 1. This line
provides the lower bound for the glass transition temperature.

\begin{figure}[ht]
\includegraphics[width=3.2in]{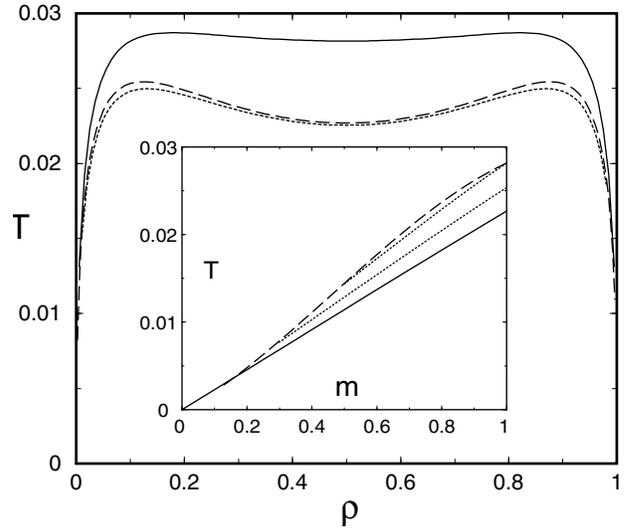}
\caption{Dynamical (solid line) and thermodynamical (dashed line) transition
temperatures as functions of density. The dotted line shows the
temperature at which the entropy of the liquid state becomes negative.
The insert shows the trajectory  $T(m)$ for thermodynamic (solid line)
and marginal (dashed line) solutions. The dotted lines represent the 
trajectories with constant configuration entropy: The upper one 
($S_{conf}=0.07286$) corresponds to the physical cooling 
process in which the system is trapped at the dynamical transition 
temperature. The lower one ($ S_{conf}=0.04$) represents a process
when a system trapped in a lower energy state.}
\end{figure}

{\it Glass state. }In the glass phase the replica symmetry is broken, we
assume that the solution has one step RSB structure and then verify that it is
indeed a stable solution. Taking $B_{\alpha ,\beta }=b_{1}\delta _{\alpha
,\beta }-b_{2}\,R_{\alpha ,\beta },A_{\alpha ,\beta }=a_{1}\delta _{\alpha
,\beta }-a_{2}\,R_{\alpha ,\beta }$ where the matrix $R$ is a block-diagonal
matrix consisting of $m\times m$ blocks with all elements equal $1$, we get
the free energy functional 
\begin{eqnarray}
\beta F_{repl}&=&\frac{m-1}{2m}[L(a_{1})-\ln (a_{1}+\frac{b_{1}}{u\beta}
)]+\frac{L(a_{1}-ma_{2})}{2m}
\nonumber \\
&-&\frac{1}{2m}\ln X-\frac{1}{m}\ln Z_{m}+\mu \bar{\rho}+\frac{1}{2}\bar{\rho}%
^{2}(b_{1}-mb_{2})  \label{frep}
\end{eqnarray}
where $X=a_{1}-ma+(b_{1}-mb_{2})/u\beta $ and 
\[
Z_{m}=\int P_{m}(y)dy,P_{m}(y)=e^{-y^{2}/2}(1+e^{h+y\sqrt{b_{2}}})^{m}/\sqrt{%
2\pi },
\]
with $h=\mu +\bar{\rho}(b_{1}-mb_{2})-b_{1}/2.$ The free energy 
(\ref{frep})  has to be taken at the saddle point with respect to $
a_{1},a_{2},b_{1},b_{2}.$ For the thermodynamic solution it should be also
at the saddle point with respect to $m,$ according to (\ref{sclon}) it means
that $S_{conf}=0.$ The numerical solution for the dependence of temperature
on $m$ at fixed density ($\bar{\rho}=0.5$) is shown in Fig 1 insert. By definition 
$0<m<1,$ thus the value of temperature for which $m=1$ determines the thermodynamic 
glass transition temperature $T_c.$ It is plotted as a function of density in Fig. 1.

{\it Stability and marginal solution. }In order to analyze stability of one step
RSB ansatz we expand the Eq.(\ref{heff}) to the second order in fluctuation of
the order parameter $\delta B$ and consider different families of
fluctuation matrices $\delta B$. This calculation is very similar to the
analysis of the stability of paramagnetic solution and Parisi solution in SK
model \cite{Almeida,DeDominicis} so we only sketch it here. We find that the
most dangerous direction in the fluctuation space corresponds to the
''replicon'' modes \cite{Almeida,DeDominicis} that are fluctuations within
diagonal blocks of $\delta B$ satisfying the conditions $(\delta
B\,R)_{\alpha ,\beta }=0$, $\delta B_{\alpha ,\alpha }=0.$ The eigenvalue
corresponding to these modes is 
\[
\Lambda =\frac{2}{(u\beta)^2 \chi^{-1}(a_{1})-(\bar{\rho}-g_{2})^{-2}}
+2(g_{2}-2g_{3}+g_{4})
\]
where $g_{k}=\int P_{m}(z)[1+\exp (-h-z\sqrt{b_{2}})]^{-k}dz/Z_{m}$ and $%
\chi (a)=\int \exp [-\varepsilon ^{2}/2](a+\varepsilon ^{2})^{-2}dx/\sqrt{%
2\pi }.$ Numerical solution shows that for any density the thermodynamic
solution is stable in the wide temperature range below $T_{c}$ but
eventually becomes unstable at low temperature. The solution corresponding
to marginally stable states is obtained by taking $\Lambda =0,$ the
corresponding $T(m)$ dependence is plotted in Fig. 1 insert. As was
explained above $m=1$ defines the dynamical glass transition temperature $%
T_{g}$, which is plotted as a function of density in Fig. 1.

{\it Physical cooling. } During the physical cooling process the system
remains trapped in a single metastable state. As was explained above, the 
physical properties of such system are described by the trajectory $m(T)$
satisfying $S_{conf}(m,T)=const$. We show two such trajectories in the 
insert of Fig. 1. The upper dotted line correspond to a typical process in which
the system got trapped in the highest energy state at the dynamical transition 
temperature         
(for this curve $S_{conf}=0.072864$), the lower one to a system 
trapped in a lower energy state corresponding to $S_{conf}=0.04$. We see that 
these trajectories cross the marginal stability line at lower temperatures, 
beyond this line the one step RSB ansatz becomes unstable. Physically, it means that a 
metastable state in which the system is trapped starts to divide into new ones.

{\it Thermodynamics of the glass transition.} General thermodynamic 
arguments show that static properties (specific heat, pressure) of the 
glass transition are controlled by the total entropy $S_{tot}=S+S_{conf}$. 
We show the temperature dependence of $S_{tot}$ and entropy of the liquid
state for density $\bar{\rho}=1/2$  in Fig 2. 
The upper dotted line represents the physical
cooling that begins from the dynamical glass transition temperature $T_{g}.$
We see that the entropy has no jump at the phase transition while its
derivative (specific heat) does. We also consider the pressure defined, as
usually, by $P=-\partial F/\partial V.$ Along the physical process $
S_{conf}=const,$ therefore $P=-\partial F_{rep}/\partial V.$ The
corresponding plots are shown in Fig. 2. As in case with the entropy, the
pressure has no jump at the phase transition, while its derivative $dP/dT$
does.

In conclusion, we formulated a simple model of the structural glass and 
solved it in the limit of $d\rightarrow \infty$ using the mean field theory. 
We constructed a full thermodynamic description of the physical cooling process 
in which the system is trapped in a single metastable state. At very low 
temperature a metastable state in which the system is trapped starts to divide 
into new states, below this temperature our method based on one step RSB ansatz 
can not be applied. From our analysis we conclude that in this system there is 
no jump in the entropy or pressure at the glass transition temperature  while 
there are jumps in their temperature derivatives. These qualitative conclusions 
are in agreement with an established phenomenological picture of the glass 
transition \cite{Kauzmann48}.

\begin{figure}[ht]
\includegraphics[width=3.5in]{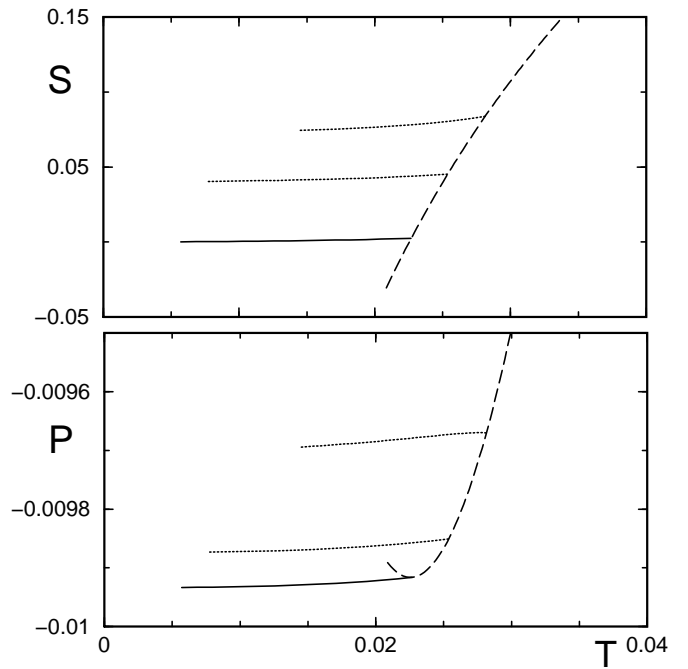}
\vspace{0.1cm}
\caption{Temperature dependence of the entropy and pressure 
in the glass phase (solid and dotted lines) and in the liquid phase (dashed lines). 
Solid lines correspond to the thermodynamic 
solution ($S_{conf}=0$), the upper dotted lines represent the 
total entropy $S_{tot}$ and pressure of the physical cooling process 
($S_{conf}=0.072864$) in which the system is trapped at the dynamical transition 
temperature, the lower dotted lines correspond to a process when a system trapped 
in a lower  energy state (corresponding to $S_{conf}=0.04$).}
\end{figure}

\end{document}